\documentclass[12pt]{article}
\usepackage{color}
\usepackage{ulem}
\usepackage{amssymb,amsmath,comment,mathtools}
\usepackage[dvipdfmx]{graphicx}
\usepackage{subfigmat}
\usepackage{braket}
\usepackage{hyperref} 	  		
\usepackage{epsfig}
\usepackage{here}
\usepackage{color}
\usepackage{bm}
\graphicspath{{./Figures/}}

\newcommand{\ba}{\begin{align}}
\newcommand{\ea}{\end{align}}

\def\nn{\nonumber}

\def\bea{\begin{eqnarray}}
\def\eea{\end{eqnarray}}
 \makeatletter
\def\alt{\mathrel{\mathpalette\gl@align<}}
\def\agt{\mathrel{\mathpalette\gl@align>}}
\def\gl@align#1#2{\lower.6ex\vbox{\baselineskip\z@skip\lineskip\z@
\ialign{$\m@th#1\hfil##\hfil$\crcr#2\crcr\sim\crcr}}} \makeatother

\usepackage{feynmp} 
\usepackage{tikz-feynman}
\tikzfeynmanset{compat=1.1.0}

\usepackage{tikz-feynhand}

\begin{document}
\begin{flushleft}
\end{flushleft}

\vspace*{1.0cm}
\begin{center}
\baselineskip 20pt 
{\Large\bf 
Confronting MSSM flat direction inflation
\\
with Planck/BICEP data
}
\vspace{1cm}

{\large 
Naoyuki Haba${}^{b,c}$, Yasuhiro Shimizu${}^{b,c}$, Yoshihiro Tanabe${}^{a,b}$, \\
and Toshifumi Yamada${}^{d}$
} \vspace{.5cm}

{\baselineskip 20pt \it
${}^{a}$Institute of Science and Engineering, Shimane University, Matsue 690-8504, Japan\\
${}^{b}$Department of Physics, Osaka Metropolitan University, Osaka 558-8585, Japan \\
${}^{c}$Nambu Yoichiro Institute of Theoretical and Experimental Physics (NITEP),
Osaka Metropolitan University, Osaka 558-8585, Japan\\
${}^{d}$Institute for Mathematical Informatics, Meiji Gakuin University, Yokohama 244-8539, Japan
}

\vspace{1.5cm} {\bf Abstract} \end{center}
\noindent

We study the scenario of inflection point inflation where a flat direction of the minimal supersymmetric standard model (MSSM) is identified with the inflaton.
Specifically, we consider in full generality the cases where a MSSM flat direction is lifted by a higher-dimensional superpotential whose dimension is $n=4,5,6,7,9$.
We confront the inflection point inflation scenarios with various $n$ with the Planck and BICEP data,
 and thereby constrain the soft SUSY breaking mass and the coefficient of the higher-dimensional operator that lifts the flat direction.

\section{Introduction}
In the beginning of the Universe, it is believed, from cosmological observations such as the measurement of the cosmic microwave background, that a rapid expansion, i.e., inflation occurred and that it was induced by a scalar field called inflaton~\cite{Guth:1980zm}.
A lot of inflation scenarios and inflaton candidates have been proposed. One of the attractive scenarios is so-called inflection point inflation~\cite{Allahverdi:2006iq}-\cite{Rodrigues:2023xqu}, since it easily satisfies the experimental constraints on the spectral index $n_s$ and the tensor-to-scalar ratio $r$.

Supersymmetric (SUSY) extension of the standard model is a viable candidate for physics above TeV scale, because SUSY can stabilize the large hierarchy between the Planck scale and the electroweak scale.
The minimal SUSY standard model (MSSM) exhibits an interesting property that certain combinations of scalar fields have vanishing triple and quartic couplings~\cite{FD1}.
Such combinations are called ``flat directions".
The flat directions are a natural candidate for the inflaton of inflection point inflation~\cite{Allahverdi:2006iq}-\cite{Allahverdi:2006we}.

In this paper, we study the scenario where a flat direction in MSSM is identified with the inflaton that induces inflection point inflation.
Specifically, we consider in full generality the cases where a MSSM flat direction is lifted by a higher-dimensional superpotential operator whose dimension is $n=4,5,6,7,9$.
These values of $n$ are based on the finding of Ref.~\cite{FD1} that the MSSM flat directions are lifted by superpotential operators with either $n=4,5,6,7$ or 9.
The authors of Ref.~\cite{FD1} have first made a complete catalog of flat directions in renormalizable MSSM, 
 and then searched for $R$-parity-conserving higher-dimensional operators that lift the flat directions.
In this way, they have found that all the flat directions are lifted by at least one of higher-dimensional operators with $n=4,5,6,7,9$.
For each $n$, we confront the corresponding inflection point inflation scenario with the Planck and BICEP data,
 and thereby constrain the soft SUSY breaking mass and the coefficient of the higher-dimensional operator that lifts the flat direction.

Previously, the study on MSSM-based inflection point inflation is centered on the case with $n=6$ as in Refs.~\cite{Allahverdi:2006iq,Allahverdi:2006we,Enqvist:2007tf,Allahverdi:2008bt,Allahverdi:2010zp,Choudhury:2013jya,Ferrantelli:2017ywq,Weymann-Despres:2023wly}, 
 because in this case TeV-scale SUSY particle masses and Planck suppression of higher dimensional operators are consistent with 
 the data on the scalar perturbation amplitude.
In the current paper, we extend the previous work to other values of $n$ and more general scales of SUSY particles masses and higher dimensional operators,
 and perform a comparative study on cases with different $n$'s.
Additionally, for the sake of generality, 
 we distinguish what is common to all inflection point inflation models and what is specific to MSSM-based ones,
 by separating the analysis into model-independent one (subsection~\ref{inflation1}) and one for MSSM flat direction (subsections~\ref{inflation2},~\ref{inflation3}).
Moreover, we employ the latest Planck and BICEP data to update the previous work.

This paper is organized as follows:
In Section~\ref{model}, we describe the model where MSSM is extended by the higher-dimensional operator that lifts a flat direction,
 and derive the potential for the flat direction that can realize inflection point inflation.
In Section~\ref{inflation}, we confront the model with the Planck and BICEP data.
First, in subsection~\ref{inflation1}, we perform a comparably model-independent analysis for inflection point inflation.
Then, in subsection~\ref{inflation2}, we focus on inflection point inflation with a MSSM flat direction,
 and derive a constraint on the soft SUSY breaking mass and the coefficient of the higher-dimensional operator for $n=5,6,7,9$.
In subsection~\ref{inflation3}, we derive a constraint for $n=4$ separately.
Section~\ref{conclusion} concludes the paper.

\section{Model}
\label{model}


The model we consider is MSSM extended by a higher-dimensional operator that lifts a flat direction.

Let us denote a canonically-normalized flat direction, which is a complex scalar field, by $\Phi$.
For example, a $udd$ flat direction is given by
\begin{equation*}
\tilde{u}^{a}_i=\dfrac{\Phi}{\sqrt{3}} ,\; \tilde{d}^b_j=\dfrac{\Phi}{\sqrt{3}} , \; \tilde{d}^c_k= \dfrac{\Phi}{\sqrt{3}},
\end{equation*}
 where $\tilde{u},\tilde{d}$ denote the scalar components of the isospin-singlet up-type and down-type quark superfields, respectively,
 subscript $i, j, k$ denote generation, and subscript $a, b, c$ are color indices.

The superpotential is given by
\begin{align}
W = W_{\rm MSSM} + \frac{\lambda}{n M_p^{n-3}}\Phi^n,
\label{superpotential}
\end{align}
 where, by abuse of notation, $\Phi$ also represents the chiral superfield that contains the scalar field $\Phi$.
Here $W_{\rm MSSM}$ is the superpotential of MSSM,
 $\lambda$ is a coupling constant that is taken to be real positive, $M_p$ is the reduced Planck mass (2.44 $\times 10^{18}$ GeV), and we assume $n\geq4$.
For example, if a $udd$ flat direction is lifted by a $R$-parity-conserving dimension-six operator as
\begin{equation*}
W = W_{\rm MSSM} + \frac{\tilde{\lambda}}{2 M_p^3}(U^c_i D^c_j D^c_k)^2,
\end{equation*}
 where $U^c,D^c$ respectively denote the chiral superfields containing $\tilde{u},\tilde{d}$,
 then we obtain
\begin{equation*}
W = W_{\rm MSSM} + \frac{\tilde{\lambda}}{3^3\cdot 2 M_p^3}\Phi^6
\end{equation*}
 and hence $\lambda$ is related to $\tilde{\lambda}$ as
\begin{equation*}
\lambda = \frac{\tilde{\lambda}}{9}.
\end{equation*}
The flat direction is also lifted by a soft SUSY breaking mass and an $A$-term proportional to the higher-dimensional operator.
As a result, the potential for the flat direction is given by
\begin{align}
V(\Phi) \ = \ m_\Phi^2 |\Phi|^2 - {\cal A} \frac{\lambda}{nM_p^{n-3}} \Phi^n - {\rm h.c.} +  \frac{\lambda^2}{M^{2(n-3)}_p}|\Phi|^{2(n-1)}
\end{align}

We write the radial and angular components of $\Phi$ as $\phi,\theta$ such that $\Phi=\phi e^{i\,\theta}/\sqrt{2}$ holds.
The potential is then recast into
\begin{align}
V(\Phi) \ = \ V(\phi,\theta) \ = \ \frac{m_\Phi^2}{2} \phi^2 - |{\cal A}| \frac{\lambda}{2^{\frac{n}{2}-1}nM_p^{n-3}} \phi^n \cos(n\theta+\theta_A)  
+  \frac{\lambda^2}{2^{n-1}M^{2(n-3)}_p} \phi^{2(n-1)},
\label{potential-pre}
\end{align}
 where $\theta_A$ denotes the phase of ${\cal A}$.
This potential is minimized for $\theta$ satisfying $\cos(n\theta+\theta_A)=1$.
Without loss of generality, we can assume that $\theta$ is stabilized at $\theta = -\theta_A/n$.
We write the deviation of $\theta$ from this minimum as $\Delta \theta = \theta+\theta_A/n$.
During slow-roll inflation, the canonically-normalized field that corresponds to $\Delta\theta$ is approximately given by $\langle \phi \rangle \Delta\theta$.
As we perform a consistency check in Appendix~A,
 the value of $\langle \phi \rangle \Delta\theta$ becomes 0 in a time much shorter than the Hubble time in inflection point inflation with $|{\cal A}|/\lambda \ll M_p$.
Therefore, in the study of inflationary dynamics, we can fix $\Delta\theta =0$ and $\cos(n\theta+\theta_A)=1$, by neglecting the dynamics of the angular field $\langle \phi \rangle \Delta\theta$.
Once we fix $\cos(n\theta+\theta_A)=1$, the potential for $\phi$, which is a real scalar field, becomes
\begin{equation}
    V(\phi)=\dfrac{m^2_\Phi}{2}\phi^2 - |\mathcal{A}|\dfrac{\lambda}{2^{\frac{n}{2}-1}n M^{n-3}_p}\phi^n + \dfrac{\lambda^2}{2^{n-1}M^{2(n-3)}_p}\phi^{2(n-1)}.
\label{potential}
\end{equation}

\section{Inflection point inflation}
\label{inflation}

\subsection{Model-independent analysis}
\label{inflation1}

Before analyzing the specific inflaton potential Eq.~(\ref{potential}), 
 we consider a generic potential $V(\phi)$ and perform a model-independent analysis for inflection point inflation.
The only assumptions we make on the generic potential $V(\phi)$ are
\begin{itemize}

\item $V(\phi)$ has a quasi-inflection point $\phi=\phi_0$ such that $V''(\phi_0)=0$ and $M_p \dfrac{V'(\phi_0)}{V(\phi_0)} <1$ hold.

\item This quasi-inflection point satisfies $\phi_0>0$, $V'(\phi_0) > 0$, $V'''(\phi_0) > 0$.

\item $M_p^3 \dfrac{V'''(\phi_0)}{V(\phi_0)} >1$ holds.
\end{itemize}

We consider the situation where $\phi$ is in the vicinity of $\phi_0$ such that $|\phi-\phi_0|\ll \phi_0$ holds.
For later convenience, we define
\begin{align}
N_0 = \frac{1}{M_p^2}\sqrt{\frac{2V(\phi_0)^2}{V'(\phi_0)V'''(\phi_0)}}.
\label{n0def}
\end{align}
The slow-roll parameters are computed as
\begin{align}
\eta(\phi) &= M_p^2\frac{V''(\phi)}{V(\phi)} \simeq M_p^2\frac{V'''(\phi_0)}{V(\phi_0)}(\phi-\phi_0),
\label{eta}\\
\varepsilon(\phi) &= \frac{M_p^2}{2}\left(\frac{V'(\phi)}{V(\phi)}\right)^2 \simeq \frac{M_p^2}{2}\left(\frac{V'(\phi_0)+\frac{1}{2}(\phi-\phi_0)^2V'''(\phi_0)}{V(\phi_0)}\right)^2
\nn\\
&=\frac{1}{8}\left(\frac{1}{M_p^3}\frac{V(\phi_0)}{V'''(\phi_0)}\right)^2 \left(\frac{4}{N_0^2}+\eta(\phi)^2\right)^2,
\label{epsilon}
\end{align}
 where in the $\simeq$ of each equation, we have made an approximation with $|\phi - \phi_0|\ll\phi_0$
 so that terms involving higher powers of $\phi-\phi_0$ are neglected.
The number of e-folds is calculated as
\begin{align}
N(\phi)&= \frac{1}{M_p^2}\int_\phi^{\phi_{\rm end}}{\rm d}\phi \frac{V(\phi)}{-V'(\phi)} \simeq \frac{1}{M_p^2}\int_\phi^{\phi_{\rm end}}{\rm d}\phi \frac{V(\phi_0)}{-V'(\phi_0)-\frac{1}{2}(\phi-\phi_0)^2V'''(\phi_0)} 
\nn\\
&= N_0 \left[ \arctan\left(\frac{N_0\ \eta(\phi)}{2}\right) - \arctan\left(\frac{N_0\ \eta(\phi_{\rm end})}{2}\right) \right].
\label{Nefolds}
\end{align}

We confront the above calculations with the cosmological observations.
We denote by $k_*$ the comoving wavenumber corresponding to the pivot scale of 0.05~Mpc$^{-1}$ at present,
 and denote by $\phi_*$ the inflaton VEV when this scale exited the horizon.
The scalar spectral index $n_s$ is given by $n_s=1-6\varepsilon(\phi_*)+2\eta(\phi_*)$, while it has been measured to be $n_s = 0.9649\pm 0.0042$~\cite{Planck:2018jri}.
The tensor-to-scalar ratio $r$ is given by $r=16\varepsilon(\phi_*)$, while it has been constrained as $r<0.032$ at 95\% CL~\cite{Tristram:2021tvh}.
Hence $\eta(\phi_*)<0$ holds at 2$\sigma$.
Then Eq.~(\ref{Nefolds}) gives $N_0 > 2N(\phi_*)/\pi$,
 where $N(\phi_*)$ corresponds to the number of e-folds since the pivot scale exited the horizon.

For $N(\phi_*) \gtrsim 20$, inserting $N_0 > 2N(\phi_*)/\pi$ into Eq.~(\ref{epsilon}), we obtain $6\varepsilon(\phi_*) \ll 2|\eta(\phi_*)|$, $n_s\simeq 1+2\eta(\phi_*)$
 and that the data of $n_s = 0.9649\pm 0.0042$ fixes $\eta(\phi_*)$ as
\begin{align}
\eta(\phi_*)=-0.018\pm 0.002.
\label{eta_fixed}
\end{align}
Additionally, we get $r=16\varepsilon(\phi_*) < 2\times 10^{-3}$ at 95\% CL and so the experimental bound on $r$ is met with ease.
We infer from Eq.~(\ref{epsilon}) that the end of inflation is determined by the relation $|\eta(\phi_{\rm end})|=1$.
Then we can fix $N_0$ in terms of $N(\phi_*)$ by inserting Eq.~(\ref{eta_fixed}) and $\eta(\phi_{\rm end})=-1$ into Eq.~(\ref{Nefolds}) and solving 
\begin{align}
N(\phi_*)=N_0\left[\arctan\left(-N_0 \cdot 0.009\pm 0.001\right)-\arctan\left(-N_0/2\right)\right]
\label{Nefolds_sol}
\end{align}
 for $N_0$.
Additionally, the inflaton VEV at the end of inflation satisfies
\begin{align}
\phi_0 - \phi_{\rm end} = \frac{V(\phi_0)}{M_p^2 V'''(\phi_0)}.
\label{phi_end}
\end{align}
The scalar power spectrum amplitude at the pivot scale $P_\zeta(k_*)$ is calculated as
\begin{align}
P_\zeta(k_*) &= \frac{V(\phi_*)}{24\pi^2M_p^4\varepsilon(\phi_*)} \simeq \frac{V(\phi_0)}{24\pi^2M_p^4\varepsilon(\phi_*)}
\nn\\
&= \frac{1}{24\pi^2}8\frac{V(\phi_0)}{M_p^4}\left(M_p^3\frac{V'''(\phi_0)}{V(\phi_0)}\right)^2\left(\frac{4}{N_0^2}+\eta(\phi_*)^2\right)^{-2},
\label{pzetakstar}
\end{align}
 while it has been measured to be $P_\zeta(k_*)=e^{3.044\pm 0.014}\times10^{-10}$~\cite{Planck:2018jri}.
This provides an independent constraint on the potential.
\\

We comment in passing that in the bona fide inflection point inflation where $V'(\phi_0)=0$ is assumed, 
 $\eta(\phi_*)$ and $N(\phi_*)$ are tightly connected as $\eta(\phi_*)=-2/N(\phi_*)$
 and hence $n_s$ is predicted to be $n_s \simeq 1+2\eta(\phi_*) = 1 - 4/N(\phi_*) \lesssim 0.93$, which is in contradiction with the Planck data. 
Therefore the bona fide inflection point inflation is excluded.
On the other hand, in our scenario where $V'(\phi_0)\neq 0$, the connection is loosened as Eq.~(\ref{Nefolds}) and the model can be consistent with the Planck data.

\subsection{Analysis for MSSM flat direction}
\label{inflation2}

Now we concentrate on the specific inflaton potential Eq.~(\ref{potential}).
In order to have a quasi-inflection point, the soft SUSY breaking parameters should satisfy
\footnote{
Our definition of $\alpha$ coincides with that of Ref.~\cite{Weymann-Despres:2023wly} in the leading order.
}
\begin{align}
m_\Phi^2 = \frac{|{\cal A}|^2}{4(n-1)}(1+\alpha),
\label{quasi-inflectionpointcondition}
\end{align}
 where $\alpha$ is a small number that parametrizes a detour from bona fide inflection point.

The point $\phi_0$ that satisfies $V''(\phi_0)=0$, $V'(\phi_0) \propto \alpha$ and $\phi_0>0$ is given by
\begin{align}
\phi_0  =  \sqrt{2}\left(\frac{|{\cal A}|M_p^{n-3}}{2\lambda(n-1)}\right)^{\frac{1}{n-2}}   \left(1-\frac{\alpha}{2(n-2)^2}\right)    + O(\alpha^2).
\label{quasiinflectionpoint}
\end{align}
The potential and its first and third derivatives at $\phi=\phi_0$ read
\begin{align}
V(\phi_0) = \frac{1}{4}\frac{(n-2)^2}{n(n-1)^2}|{\cal A}|^2\left(\frac{|{\cal A}|M_p^{n-3}}{2\lambda(n-1)}\right)^{\frac{2}{n-2}} + O(\alpha),
\label{v0}\\
V'(\phi_0) = \alpha\frac{1}{2\sqrt{2}(n-1)}|{\cal A}|^2\left(\frac{|{\cal A}|M_p^{n-3}}{2\lambda(n-1)}\right)^{\frac{1}{n-2}} + O(\alpha^2),
\\
V'''(\phi_0) = \frac{(n-2)^2}{2\sqrt{2}(n-1)}|{\cal A}|^2\left(\frac{|{\cal A}|M_p^{n-3}}{2\lambda(n-1)}\right)^{-\frac{1}{n-2}} + O(\alpha).
\label{v03}
\end{align}

We choose $\alpha>0$ so that $V'(\phi_0)>0$ holds.
$\phi_0$ should satisfy $M_p V'(\phi_0)/V(\phi_0) <1$ in order to be qualified as a quasi-inflection point.
This condition is translated into the following bound on $\alpha$:
\begin{align}
M_p \frac{V'(\phi_0)}{V(\phi_0)} <1 \ \Rightarrow \ \alpha < \frac{(n-2)^2}{\sqrt{2}n(n-1)}\left(\frac{|{\cal A}|}{2\lambda M_p(n-1)}\right)^{\frac{1}{n-2}}.
\label{alphabound1}
\end{align}

When $|{\cal A}|/\lambda \ll M_p$, we get $\phi_0 \ll M_p$ (the small-field inflation is realized)
 and we obtain $M_p^3 V'''(\phi_0)/V(\phi_0) >1$.
If further Eq.~(\ref{alphabound1}) holds, the assumptions in subsection~\ref{inflation1} are all satisfied
 and the results of subsection~\ref{inflation1} can be carried over.

Next we derive the expression for the scalar power spectrum amplitude $P_\zeta(k_*)$.
Inserting Eqs.~(\ref{v0}),(\ref{v03}) into Eq.~(\ref{pzetakstar}), we get
\begin{align}
P_\zeta(k_*) &= \frac{1}{6\pi^2}n(n-2)^2\left(\frac{|{\cal A}|}{M_p}\right)^{2-\frac{4}{n-2}}\left(2\lambda(n-1)\right)^{\frac{4}{n-2}}\left(\frac{4}{N_0^2}+\eta(\phi_*)^2\right)^{-2}.
\label{scalaramplitude}
\end{align}
Here $N_0$ is determined from the number of e-folds since the pivot scale exited the horizon $N(\phi_*)$ through Eq.~(\ref{Nefolds_sol}),
 and $N(\phi_*)$ depends on the reheating temperature, $T_R$, as reviewed in Appendix~B.

$T_R$ is estimated in the following manner:
The inflaton $\phi$ decays into SM and MSSM particles and reheats the Universe.
In this paper, we restrict our study to the case where the Hubble rate at the end of inflation $H(\phi_{\rm end})$ is on the same order or smaller than the decay rate of the inflaton.
As we will see in Fig.~\ref{result2}, in most of the interesting cases, $H(\phi_{\rm end}) \lesssim 10^{-2} m_\Phi$ does hold and hence 
 this restriction does not destroy the generality of the study.
When $H(\phi_{\rm end})$ is on the same order or smaller than the inflaton decay rate, cosmic expansion since the end of inflation until the time when all the inflatons have decayed,
 is negligible.
It follows that no matter how the inflaton reheats the Universe, the scale factors at the end of inflation and at the reheating are approximately the same.
Also, radiation energy density when all the inflatons have decayed can be approximated by vaccum energy density at the end of inflation, due to energy conservation.
Namely, the scale factor and $T_R$ satisfy the following relations:
\begin{align}
a(t_{\rm rh})&\simeq a(t_{\rm end}),
\label{case1-1}\\
\frac{\pi^2}{30}g_{\rm eff} T_R^4 &\simeq 3M_p^2 H(\phi_{\rm end})^2,
\label{case1-2}
\end{align}
 where $g_{\rm eff}$ is the effective relativistic degree of freedom at temperature $T=T_R$, 
 and $a(t_{\rm end}),a(t_{\rm rh})$ are respectively the scale factor at the end of inflation and at the reheating temperature.

\begin{figure}[H]
    \includegraphics[width=10cm]{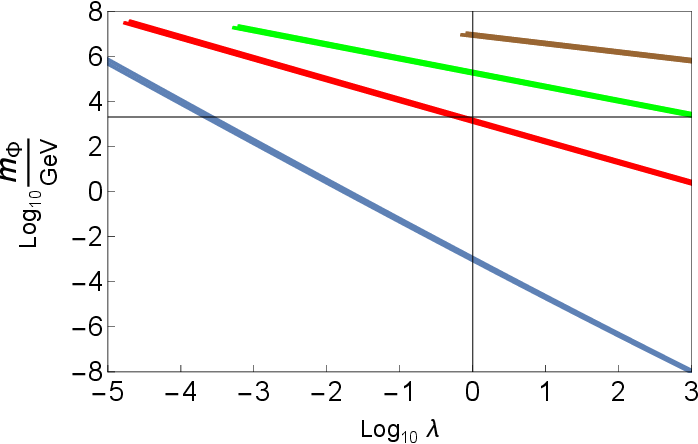}
  \caption{
  Prediction on the relation between $\lambda$ and soft SUSY breaking mass $m_{\Phi}\simeq |{\cal A}|/(2\sqrt{n-1})$.
  The blue, red, green, brown lines correspond to $n=5,6,7,9$, respectively.
  The horizontal line is the line of $m_{\Phi}=2000$~GeV, roughly corresponding to the lower bounds of SUSY particle masses at the LHC.
  The colored lines are discontinued when $H(\phi_{\rm end})$ surpasses $10^{-2} m_\Phi$, because we focus on the case with $H(\phi_{\rm end}) \leq 10^{-2} m_\Phi$ so that $H(\phi_{\rm end})$ is on the same order or smaller than the inflaton decay rate.
    }
  \label{result1}
\end{figure}

\begin{figure}[H]
    \includegraphics[width=10cm]{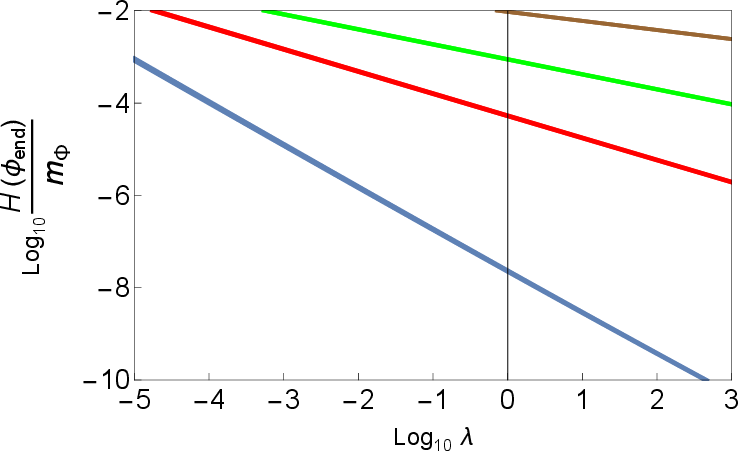}
  \caption{
  Prediction on the relation between $\lambda$ and the ratio of the Hubble rate at the end of inflation over the soft SUSY breaking mass $H(\phi_{\rm end})/m_{\Phi}$.
 By comparing this figure with Fig.~\ref{result1}, one verifies that the colored lines are discontinued when $H(\phi_{\rm end})$ surpasses $10^{-2} m_\Phi$.
    }
  \label{result2}
\end{figure}

To evaluate $H(\phi_{\rm end})$, we note that from Eqs.~(\ref{v0}),(\ref{v03}) and Eq.~(\ref{phi_end}), we have $|\phi_{\rm end}-\phi_0|\ll \phi_0$ and thus 
 the Hubble rate during inflation, $H_{\rm inf}$, and that at the end of inflation $H(\phi_{\rm end})$ are approximately given by
\begin{align}
H(\phi_{\rm end}) &\simeq H_{\rm inf} \simeq \sqrt{\frac{V(\phi_0)}{3M_p^2}}
=\frac{n-2}{2\sqrt{3n}(n-1)}|{\cal A}|\left(\frac{|{\cal A}|}{2\lambda M_p(n-1)}\right)^{\frac{1}{n-2}}.
\label{h_end}
\end{align}
We insert the above expression into Eq.~(\ref{case1-2}) and determine $T_R$, 
 and then use Eq.~(\ref{nefoldspivot}), $T_R$, and $H_{\rm inf}$ given above, to express $N(\phi_*)$ in terms of $n,|{\cal A}|,\lambda$. 
For the value of $g_{\rm eff}$, we use the one with the MSSM particle content, $g_{\rm eff}=915/4$.
Finally, we express $N_0$ in terms of $n,|{\cal A}|,\lambda$ using Eq.~(\ref{Nefolds_sol}), insert it into Eq.~(\ref{scalaramplitude}),
 and compare it with the data of $P_\zeta(k_*)=e^{3.044\pm 0.014}\times10^{-10}$, to constrain $|{\cal A}|,\lambda$ for each $n$.
The result is presented in Fig.~\ref{result1} for $n=5,6,7,9$. The case with $n=4$ is special and will be discussed in the next subsection.~\footnote{In the analyses for $n=4,5,6,7,9$, we have confirmed $N(\phi_*)\gtrsim 20$ and thereby justified the assumption in deriving Eq.~(\ref{eta_fixed}).
}

In Fig.~\ref{result1}, we show the prediction on the relation between $\lambda$ and soft SUSY breaking mass $m_{\Phi}\simeq |{\cal A}|/(2\sqrt{n-1})$~\footnote{See Eq.~(\ref{quasi-inflectionpointcondition}).},
 where the blue, red, green, brown lines correspond to $n=5,6,7,9$, respectively.
The region with $\lambda>1$ should be regarded as the situation where some novel physics below the Planck scale is responsible for lifting the MSSM flat directions.
For each $n$, there are actually four minute lines that correspond to the $2\sigma$ upper and lower edges of the data of $\eta(\phi_*)$ and $P_\zeta(k_*)$ given by Eq.~(\ref{eta_fixed}) and 
 $P_\zeta(k_*)=e^{3.044\pm 0.014}\times10^{-10}$,
 but the four lines are almost degenerate and indistinguishable.
The horizontal line is the line of $m_{\Phi}=2000$~GeV, which is about the lower bounds of SUSY particle masses obtained at the LHC.
This means that the colored lines below this horizontal line are in tension with the LHC results.
The colored lines are discontinued when $H(\phi_{\rm end})$ surpasses $10^{-2} m_\Phi$,
 because we focus on the case with $H(\phi_{\rm end}) \leq 10^{-2} m_\Phi$ so that $H(\phi_{\rm end})$ is on the same order or smaller than the inflaton decay rate.
For reference, the relation between $\lambda$ and $H(\phi_{\rm end})/m_{\Phi}$ is presented in Fig.~\ref{result2}.
\noindent
It is noteworthy that for $n=6$ and $\lambda=1$, $m_{\Phi}$ is predicted to be around the bound at the LHC.
This fact gives a strong phenomenological motivation for the MSSM inflection point inflation scenario with $n=6$.

We comment on the resolution of the horizon problem (homogeneity problem).
We have numerically found that $(2\pi/3)N_0 - N(\phi_*) > 6$ holds on the colored lines of Fig.~\ref{result1}.
This means that if we set the inflaton VEV at the onset of inflation, $\phi_{\rm onset}$, in the range corresponding to
 $1 > \eta(\phi_{\rm onset}) > 2/(\sqrt{3}N_0)$
 \footnote{
 Recall Eq.~(\ref{Nefolds}) for the relation between $\eta(\phi)$ and the number of e-folds.
 },
 then the total number of e-folds $N_{\rm total}$ is sufficiently large that
 $N_{\rm total} - N(\phi_*) > 6$ holds.
Then the horizon problem is solved, as reviewed in Appendix~B.

\begin{figure}[H]
    \includegraphics[width=10cm]{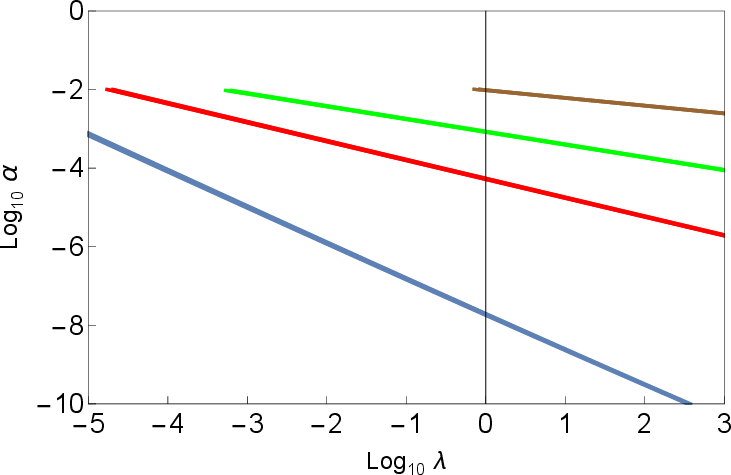}
  \caption{
  Prediction on the relation between $\lambda$ and the upper bound of the detour from bona fide inflection point $\alpha$ found in Eq.~(\ref{alphabound1}).
  }
  \label{resultalpha}
\end{figure}

We comment on the degree of fine-tuning of the model parameters.
There are two fine-tuned parameters: The detour from bona fide inflection point $\alpha$,
 and the difference between the inflaton VEV at the onset of inflation $\phi_{\rm onset}$ and the quasi-inflection point $\phi_0$.
For the fine-tuning of $\alpha$, we show in Fig.~\ref{resultalpha} the upper bound of $\alpha$ found in Eq.~(\ref{alphabound1}).
This figure informs us of the degree of fine-tuning of a quasi-inflection point necessary for successful inflection point inflation.
\noindent
For the difference between $\phi_{\rm onset}$ and $\phi_0$,
 it should be fine-tuned such that $1 > \eta(\phi_{\rm onset}) > 2/(\sqrt{3}N_0)$ holds~\footnote{ 
Note that for $|\eta| < 1$, the slow-roll condition for $\varepsilon$ is automatically satisfied thanks to Eq.~(\ref{epsilon}).
}, 
 to solve the horizon problem.
We have numerically found $2/(\sqrt{3}N_0) < 0.1$, and so
 the fine-tuning for setting $1 > \eta(\phi_{\rm onset})$ is more important than that for setting $\eta(\phi_{\rm onset}) > 2/(\sqrt{3}N_0)$.
Thus, the degree of fine-tuning of the difference between $\phi_{\rm onset}$ and $\phi_0$ is measured by the following quantity:
\begin{align}
1 > \eta(\phi_{\rm onset}) \ \Leftrightarrow \ \frac{\phi_{\rm onset}-\phi_0}{\phi_0} < \frac{1}{2n(n-1)}\left(\frac{|{\cal A}|}{2\lambda(n-1)M_p}\right)^{\frac{2}{n-2}}.
\label{finetuningonset}
\end{align}
Interestingly, Eq.~(\ref{finetuningonset}) can be rewritten by using Eq.~(\ref{alphabound1}) as
\begin{align}
\frac{\phi_{\rm onset}-\phi_0}{\phi_0} < \frac{n(n-1)}{(n-2)^4} \left({\rm upper \ bound \ of} \ \alpha\right)^2.
\label{finetuningonset2}
\end{align}
Hence the degree of fine-tuning of $(\phi_{\rm onset}-\phi_0)/\phi_0$ is about the square of that of $\alpha$.

\begin{figure}[H]
    \includegraphics[width=10cm]{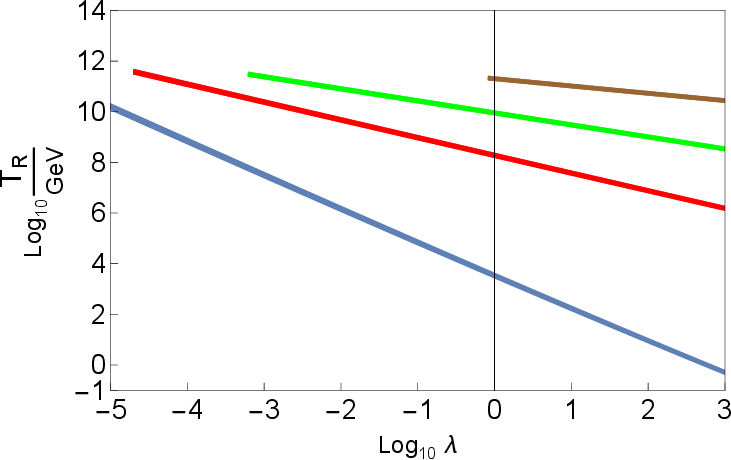}
  \caption{
  Prediction on the relation between $\lambda$ and the reheating temperature $T_R$ calculated from Eqs.~(\ref{case1-2}),(\ref{h_end}).
  }
  \label{resultTr}
\end{figure}

We comment on the reheating temperature.
In the current analysis, the reheating temperature $T_R$ is calculated from Eqs.~(\ref{case1-2}),(\ref{h_end}).
The value of $T_R$ thus obtained is plotted in Fig.~\ref{resultTr}.
\noindent
We see that the reheating temperature $T_R$ is always above 100~MeV and hence big bang nucleosynthesis works successfully~\cite{Workman:2022ynf}.
As for baryogenesis, for $n=5$ and $\lambda \gtrsim 10$, the reheating temperature $T_R$ is below the sphaleron temperature.
Even in this case, baryogenesis can be achieved through a non-thermal baryogenesis scenario where the flat direction possesses non-zero baryon number and its decay at the reheating generates baryon asymmetry of the Universe
\footnote{
For reference, non-thermal baryogenesis in MSSM-based inflection point inflation with $n=6$ and $\lambda \sim 1$ is discussed in Ref.~\cite{Haba:2024she}.
}.
$T_R$ can be bounded from above by the gravitino problem. However, this bound depends on the SUSY breaking mechanism and so it is beyond the scope of the current study.

Finally, we make a brief comment on the tensor-to-scalar ratio $r$.
On the colored lines of Fig.~\ref{result1}, we have found that $r$ is predicted to be below $10^{-13}$.
Hence, it is hopeless to discover the tensor perturbations in future experiments for $n=5,6,7,9$.

\subsection{Analysis for the case with $n=4$}
\label{inflation3}

\begin{figure}[H]
    \includegraphics[width=10cm]{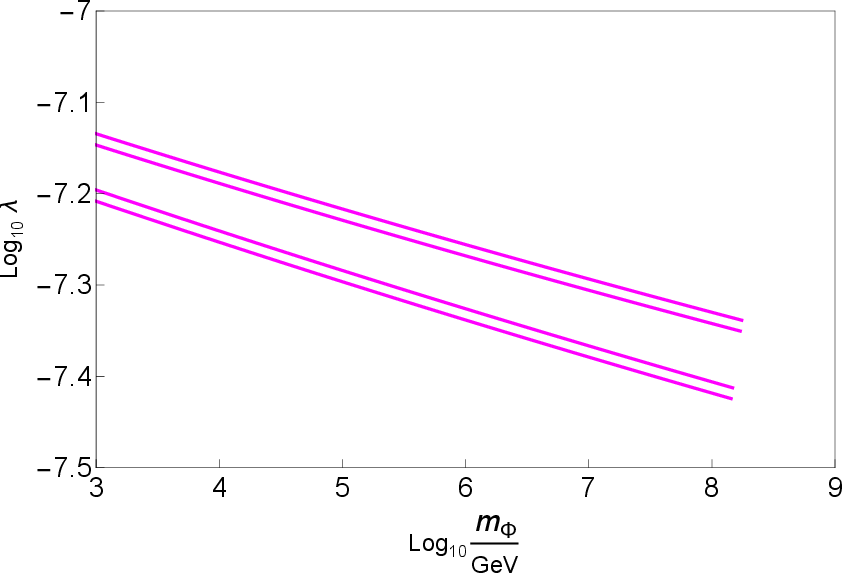}
  \caption{
  Prediction on the relation between soft SUSY breaking mass $m_{\Phi}\simeq |{\cal A}|/(2\sqrt{n-1})$ and $\lambda$ for $n=4$.
  The four lines correspond to the $2\sigma$ upper and lower edges of the data of $\eta(\phi_*)$ and $P_\zeta(k_*)$.
  The lines are discontinued when $H(\phi_{\rm end})$ surpasses $10^{-2} m_\Phi$, because we focus on the case with $H(\phi_{\rm end}) \leq 10^{-2} m_\Phi$ so that $H(\phi_{\rm end})$ is on the same order or smaller than the inflaton decay rate.
      }
  \label{n4-1}
\end{figure}

\begin{figure}[H]
    \includegraphics[width=10cm]{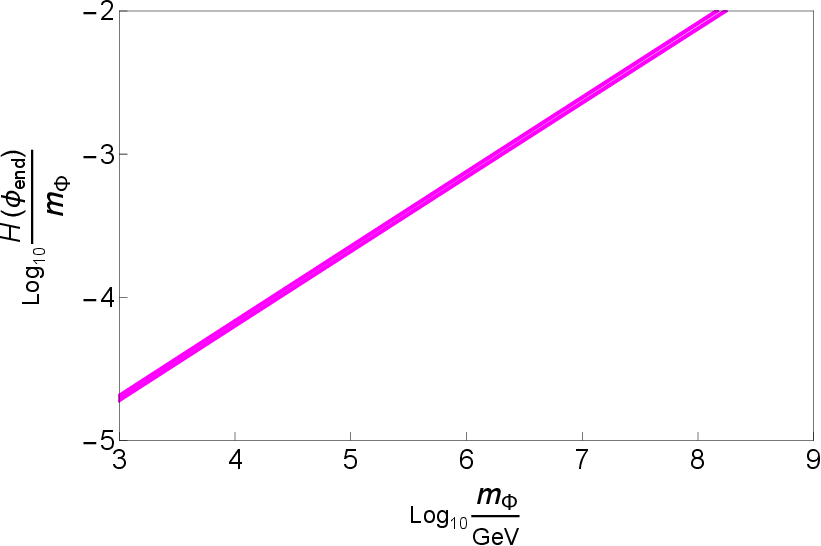}
  \caption{
  Prediction on the relation between soft SUSY breaking mass $m_{\Phi}$ and the ratio of the Hubble rate at the end of inflation over the soft SUSY breaking mass $H(\phi_{\rm end})/m_{\Phi}$.
 By comparing this figure with Fig.~\ref{n4-1}, one verifies that the lines are discontinued when $H(\phi_{\rm end})$ surpasses $10^{-2} m_\Phi$.
      }
  \label{n4-2}
\end{figure}

For $n=4$, the scalar power spectrum amplitude $P_\zeta(k_*)$ Eq.~(\ref{scalaramplitude}) depends on $|{\cal A}|$ only through $N_0,N(\phi_*),H_{\rm inf},T_R$ as Eqs.~(\ref{Nefolds_sol}),(\ref{case1-2}),(\ref{h_end}),(\ref{nefoldspivot}) and thus the dependence is very mild.
Therefore, we present the results with respect to soft SUSY breaking mass $m_{\Phi}\simeq |{\cal A}|/(2\sqrt{n-1})$.
Figs.~\ref{n4-1},\ref{n4-2} respectively show the relation between $m_{\Phi}$ and $\lambda$, and that between $m_{\Phi}$ and $H(\phi_{\rm end})/m_{\Phi}$.
Again, there are four minute lines that correspond to the $2\sigma$ upper and lower edges of the data of $\eta(\phi_*)$ and $P_\zeta(k_*)$, which are almost degenerate in the last two figures.
The lines are discontinued when $H(\phi_{\rm end})$ surpasses $10^{-2} m_\Phi$, because we focus on the case with $H(\phi_{\rm end}) \leq 10^{-2} m_\Phi$.
We see from Fig.~\ref{n4-1} that $\lambda$ should be much smaller than 1, namely, the higher-dimensional operator should be suppressed by a scale much larger than the Planck scale,
 in order for the inflection point inflation scenario with $n=4$ to be compatible with the experimental data.
Therefore, the phenomenological motivation for the scenario with $n=4$ is relatively weak.

We comment on the resolution of the horizon problem.
We have numerically found that $(\pi/2)N_0 - N(\phi_*) > 6$ holds on the colored lines of Fig.~\ref{n4-1}.
This means that if we set $\phi_{\rm onset}$ in the range corresponding to
 $1 > \eta(\phi_{\rm onset}) > 0$,
 then the total number of e-folds $N_{\rm total}$ is sufficiently large that
 $N_{\rm total} - N(\phi_*) > 6$ holds and the horizon problem is solved.

\begin{figure}[H]
    \includegraphics[width=10cm]{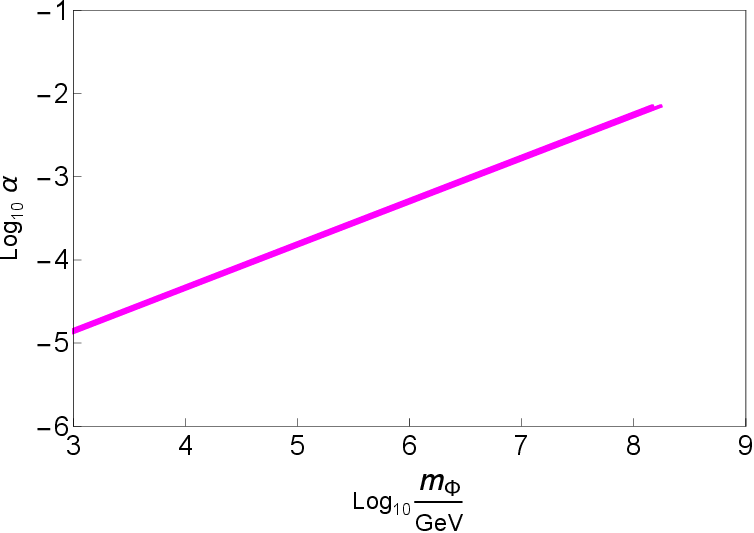}
  \caption{
  Prediction on the relation between soft SUSY breaking mass $m_{\Phi}$ and the upper bound of the detour from bona fide inflection point $\alpha$ found in Eq.~(\ref{alphabound1}).
  }
  \label{n4-3}
\end{figure}

We comment on the degree of fine-tuning of the detour from bona fide inflection point $\alpha$,
 and of the difference between the inflaton VEV at the onset of inflation $\phi_{\rm onset}$ and the quasi-inflection point $\phi_0$.
For $\alpha$, we plot in Fig.~\ref{n4-3} the relation between $m_{\Phi}$ and the upper bound of $\alpha$ found in Eq.~(\ref{alphabound1}).
This figure clarifies the degree of fine-tuning of $\alpha$ necessary for the model to be viable.
\noindent
For the difference between $\phi_{\rm onset}$ and $\phi_0$, it should be fine-tuned such that $1 > \eta(\phi_{\rm onset}) > 0$ holds, to solve the horizon problem.
Therefore, its degree of fine-tuning is measured by the same quantity as Eq.~(\ref{finetuningonset}), which can be recast into Eq.~(\ref{finetuningonset2}).
Thus the degree of fine-tuning of $(\phi_{\rm onset}-\phi_0)/\phi_0$ is about the square of that of $\alpha$, as with the case of $n=5,6,7,9$.

\begin{figure}[H]
    \includegraphics[width=10cm]{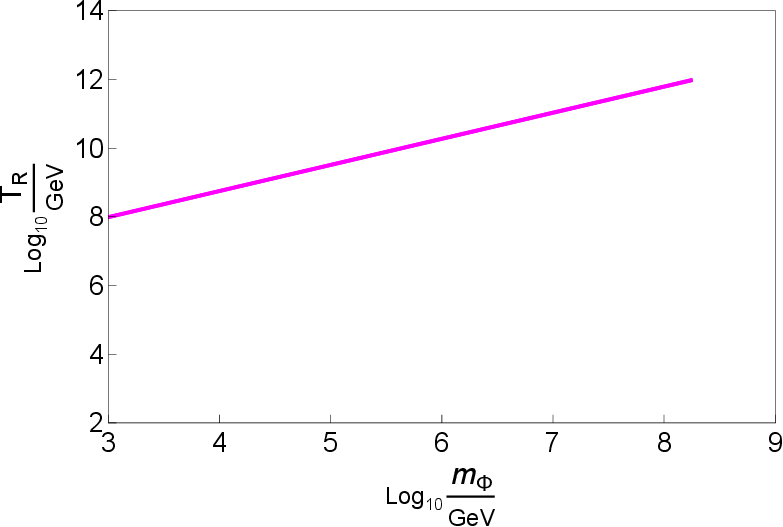}
  \caption{
  Prediction on the relation between $\lambda$ and the reheating temperature $T_R$ calculated from Eqs.~(\ref{case1-2}),(\ref{h_end}).
  }
  \label{resultTr4}
\end{figure}

We comment on the reheating temperature.
The value of $T_R$ calculated from Eqs.~(\ref{case1-2}),(\ref{h_end}) is plotted in Fig.~\ref{resultTr4}.
\noindent
We see that the reheating temperature $T_R$ is always above 100~MeV and hence big bang nucleosynthesis works successfully~\cite{Workman:2022ynf}.
Also, it is above the sphaleron temperature and thus various baryogenesis mechanisms such as leptogenesis and electroweak baryogenesis
 can be implemented.
The constraint on $T_R$ from the gravitino problem is beyond the scope of the current study, since it is dependent on the SUSY breaking mechanism.

Finally, we comment that we have found that the tensor-to-scalar ratio $r$ is predicted to be below $10^{-13}$ and thus the discovery of the tensor perturbations is impossible.

\section{Conclusion}
\label{conclusion}

We have studied inflection point inflation with a MSSM flat direction.
We have confronted the inflection point inflation scenarios with $n=4,5,6,7,9$ ($n$ is the dimension of the higher-dimensional superpotential that lifts a MSSM flat direction)
 with the Planck and BICEP data, and thereby constrained the soft SUSY breaking mass and the coefficient of the higher-dimensional operator.
We have found that the scenario with $n=6$ has a strong phenomenological motivation, since it is compatible with the experimental data with $m_\Phi \sim 2$~TeV, near the bound at the LHC, and with $\lambda \sim 1$, a natural value for higher-dimensional operators.

\section*{Acknowledgment}
This work is partially supported by Scientific Grants by the Ministry of Education, Culture,
Sports, Science and Technology of Japan, No. 23H03392 (NH) and No. 19K147101 (TY).

\section*{Appendix A}
We study the dynamics of the canonically-normalized field $\langle \phi \rangle \Delta\theta$ that corresponds to the deviation of $\theta$ from its minimum in Eq.~(\ref{potential-pre}),
 in order to check the consistency of neglecting its dynamics in the study of inflationary dynamics in inflection point inflation with $|{\cal A}|/\lambda \ll M_p$.

The mass term for the field $\langle \phi \rangle \Delta\theta$ is
\begin{align}
V(\phi,\theta) \ \supset \ \frac{1}{2}|{\cal A}| \frac{n\,\lambda}{2^{\frac{n}{2}-1}M_p^{n-3}}\langle\phi\rangle^{n-2}\left(\langle \phi \rangle \Delta\theta\right)^2.
\end{align}
In inflection point inflation, $\langle \phi\rangle$ is approximated by the field value at the quasi-inflection point $\phi_0$ Eq.~(\ref{quasiinflectionpoint}).
On the other hand, we will see that the Hubble rate $H_{\rm inf}$ during inflection point inflation is given by Eq.~(\ref{h_end}).
Comparison of the mass term and the Hubble rate gives
\begin{align}
\frac{|{\cal A}| \frac{n\,\lambda}{2^{\frac{n}{2}-1}M_p^{n-3}}\langle\phi\rangle^{n-2}}{H_{\rm inf}^2} = 
\frac{|{\cal A}| \frac{n\,\lambda}{2^{\frac{n}{2}-1}M_p^{n-3}}\phi_0^{n-2}}
{\frac{(n-2)^2}{12n(n-1)^2}{|\cal A}|^2 \left(\frac{ |{\cal A}| }{2\lambda M_p (n-1)}\right)^{\frac{2}{n-2}}}
=\frac{6n^2(n-1)}{(n-2)^2}\left(\frac{|A|}{2\lambda M_p(n-1)}\right)^{-\frac{2}{n-2}}
\end{align}
When $|{\cal A}|/\lambda \ll M_p$, the above ratio is much smaller than 1. Therefore, the value of $\langle \phi \rangle \Delta\theta$ becomes 0 in a time much shorter than the Hubble time.

\section*{Appendix B}

In order to solve the horizon problem (homogeneity problem), the Hubble distance during inflation should be larger than that at present.
This condition is translated into the following bound on the total number of e-folds $N_{\rm total}$ :
 \begin{align}
 \frac{1}{H_{\rm inf}} e^{N_{\rm total}} \frac{a(t_{\rm rh})}{a(t_{\rm end})}\frac{a_0}{a(t_{\rm rh})} \ > \ \frac{1}{H_0},
 \end{align}
 where $H_{\rm inf}$ is the Hubble rate during inflation, $t_{\rm end},t_{\rm rh}$ respectively denote
 the time at the end of inflation and at the reheating, $a_0$ is the scale factor at present, and $H_0=67$~km/s/Mpc is the Hubble rate at present~\cite{Workman:2022ynf}.
If the entropy is conserved from $t=t_{\rm rh}$ to the present, we have
 $a_0/a(t_{\rm rh})=(g_{\rm eff}T_R^3/g_{S,{\rm eff},0}T_0^3)^{1/3}$, where $T_R$ is the reheating temperature,
 $g_{\rm eff}$ is the effective relativistic degree of freedom at the reheating,
 $T_0=2.73$~K is the CMB temperature at present, and $g_{S,{\rm eff,now}}=43/11$ is the effective entropy degree of freedom at present.
By inserting the above values, the condition for solving the horizon problem is recast into
\begin{align}
N_{\rm total} > 68 - \log\frac{a(t_{\rm rh})}{a(t_{\rm end})} + \log\frac{H_{\rm inf}}{1~{\rm GeV}} - \log\frac{T_R}{1~{\rm GeV}} - \frac{1}{3}\log g_{\rm eff}.
\label{nefoldstotal}
\end{align}
Here $a(t_{\rm rh})/a(t_{\rm end})$ depends on details of the reheating process.


The number of e-folds since the comoving scale $k_*$ exited the horizon until the end of inflation $N(\phi_*)$ satisfies
\begin{align}
H_{\rm inf} \simeq H(\phi_*)=\frac{k_*}{a(t_*)} = \frac{k_*}{a_0}\frac{a_0}{a(t_{\rm rh})}\frac{a(t_{\rm rh})}{a(t_{\rm end})}e^{N(\phi_*)}.
\end{align}
For $k_*/a_0 = 0.05$~Mpc$^{-1}$, we get
\begin{align}
N(\phi_*) = 62 - \log\frac{a(t_{\rm rh})}{a(t_{\rm end})} + \log\frac{H_{\rm inf}}{1~{\rm GeV}} - \log\frac{T_R}{1~{\rm GeV}} - \frac{1}{3}\log g_{\rm eff}.
\label{nefoldspivot}
\end{align}
For this $k_*$, the condition for solving the horizon problem is re-expressed as
\begin{align}
N_{\rm total} > 6+N(\phi_*).
\label{horizonproblem}
\end{align}


\end{document}